\documentclass[]{article}
\usepackage[margin=0.5in]{geometry}
\usepackage{braket}
\usepackage{mathtools}
\usepackage{comment}
\usepackage{hyperref}
\usepackage{graphicx,subfigure}
\usepackage{amsfonts}
\usepackage{float}

\usepackage[toc,page]{appendix}

\usepackage{authblk}

\title{Variational quantum circuits to prepare low energy symmetry states}
\author[1]{Raja Selvarajan}
\author[2]{Manas Sajjan}
\author[1,2,3,*]{Sabre Kais}

\affil[1]{Purdue University, Department of Physics and Astronomy, West Lafayette, IN 47907, USA}
\affil[2]{Purdue University, Department of Chemistry, West Lafayette, IN 47907, USA}
\affil[3]{Purdue Quantum Science and Engineering Institute, West Lafayette, IN 47907, USA}
\affil[*]{Corresponding author: Sabre Kais kais@purdue.edu}

\begin{document}

\maketitle

\begin{abstract}
We explore how to build quantum circuits that compute the lowest energy state corresponding to a given Hamiltonian within a Symmetry subspace by explicitly encoding it into the circuit. We create an explicit unitary and a variationally trained unitary that maps any vector output by ansatz $A(\vec{\alpha})$ from a defined subspace to a vector in the symmetry space. The parameters are trained varitionally to minimize the energy thus keeping the output within the labelled symmetry value. The method was tested for a spin XXZ hamiltonian using rotation and reflection symmetry and $H_2$ hamiltonian within $S_z=0$ subspace using $S^2$ symmetry. We have found the variationally trained unitary surprisingly giving very good results with very low depth circuits and can thus be used to prepare symmetry states within near term quantum computers.
\end{abstract}

\section{Introduction}

One of the most important problems quantum computers have been envisioned to solve is the simulation of Hamiltonian dynamics and computation of ground state energies. Phase estimation algorithm \cite{QPE,2000quantum} despite solving this problem requires circuits that run deep, something that cannot be afforded for within the NISQ \cite{nisq} era. Alternative algorithms based on hybrid models that make use of variational methods, for instance variational quantum eigensolvers (VQE) \cite{vqe} and quantum imaginary time evolution \cite{qite} have been found to be more resilient to noisy quantum devices \cite{PhysRevX.6.031007}. The strength of a variational method depends a lot on the variational ansatz that has been used for the simulation of the state. Unitary coupled clusters \cite{gucc} decomposed using trotterization \cite{trotter}, RBM  ansatz \cite{RBM} based models and using hardware efficient gates to create generalized layered models \cite{param_hard} are some of the circuit designs that have been studied under variational methods. 

Solving for the low energy states that lie within a Symmetry subspace ,i.e, labelled by a specified symmetry value, forms a sub class within the generalized constrained optimization problems, where the idea is to minimize a cost function subject to a given set of constraints. In case where the operators of the cost function commute with that of the constraints, one could straightforwardly penalize  the cost function with an additional term that captures the error in symmetry value \cite{constrained}. Alternatively one could design circuits that variationally only explore the symmetry subspace by defining the circuit using well defined structured features. These methods explore a smaller Hilbert space for optimization and are likely to converge faster. Barkoutsos et al \cite{Barkoutsos} used a particle conserving gate alongside a particle hole conserving representation to produce ground states of simple molecular systems.  Gard et al introduced a systematic way of preserving symmetry subspaces for particle number, total spin, spin projection and time reversal  \cite{gard}. 

Unlike previous studies which aimed to provide algorithms for very specific symmetries, we would like to demonstrate the efficacy of two new techniques that can tackle any symmetry generically. We further compute the ground state energy of a given Hamiltonian constrained to lie within the specified symmetry subspace, i.e, the state is a eigenvector of the chosen symmetry operator with a user-defined eigenvalue. 

The organization of the paper is as follows. In Section 2 we illustrate in detail the underlying theoretical framework of both the methods. In Section 3 we discuss the results using Heisenberg XXZ-Spin Hamiltonian and choose two symmetry operators pertinent to the system. We also explore a real molecular system ($H_2$) to filter singlet states of total spin angular momentum squared ($S^2$) as the corresponding symmetry operator. We conclude thereafter in Section 5 with a brief discussion of possible future extensions.

\section{Method}

Hybrid variational quantum algorithms have been extensively studied in the context of solving unconstrained  \cite{nam2019groundstate, PhysRevX.8.011021,mccaskey2019quantum,quantum_ucc} and even constrained optimization problems \cite{PhysRevResearch.3.013197, QML_2021,constrained}. The primary workhorse of such methods revolve around iteratively minimizing a cost function through a usual gradient-based optimization scheme to tweak the parameters of the quantum circuit subsequently. 
The gradients of the cost-function are typically computed directly from the quantum circuit \cite{2019_gradients}, for instance using a parameter shift method \cite{2018_gradients}, while the succeeding parameter updates proceeds classically. Updates are expressed as expectation values of output states against some hermitian operator that can be re-expressed as a pauli string sum and computed independently using measurements in the basis that diagonalizes the operator or using the hadamard test \cite{hadamard_test}. The output state is typically representative of an ansatz that is expressive enough to explore a sizeable portion of the whole Hilbert space of dimension that scales as $2^n$ where $n$ is the number of qubits used in the variational method. The ansatz so chosen is generic and oblivious to the symmetry eigensector being sought.

In this work, we explore an alternative route. The cornerstone of our technique lies in its ability to confine the state-ansatz apriori to a particular eigenspace of the chosen symmetry operator even before the optimization for minimal energy is  attempted. This is attained by building into the ansatz a variational state that selectively explores the symmetry-subspace of interest only.

We now define the problem formally which shall be attempted to be solved in this work. Let us consider a system characterized by a hamiltonian $H$ and let the complete set of symmetry operators for this system be a set $P=\{O_k\}_{k=1}^N$ wherein $[H, O_m]=0 \:\:\forall\:\: O_m \in P$. Thus the set $H \bigcup P$ completely characterizes the state-space of the physical system.
We select an operator $O_k \:\:\in P$ based on a user-defined choice and find the target state $|\psi^*\rangle$ through the following optimization scheme.
\begin{eqnarray}\label{gen_opt_sch}
|\psi^*\rangle &=& \underset{\psi (\vec{\theta}) \:\in\: \Omega}{argmin_{\psi(\vec{\theta})}} \langle \psi (\vec{\theta}) | H | \psi(\vec{\theta}) \rangle \nonumber \\
such\:that \:\:\:\: \Omega &=& \{|x\rangle |\:\: O_k |x\rangle = S |x\rangle, \:\:\forall\:\: |x\rangle \in \mathbb{C}^{2^n}\} 
\end{eqnarray}
where $S$ is the user-specified eigenvalue of $O_k$ which labels the desired eigenspace and $\vec{\theta}$ are the variational parameters.  Without loss of generality one can envision that $|\Omega| =k$ i.e. the subspace is $k$-dimensional. 

\begin{figure}[H]
\centering
\includegraphics[width=0.75\textwidth]{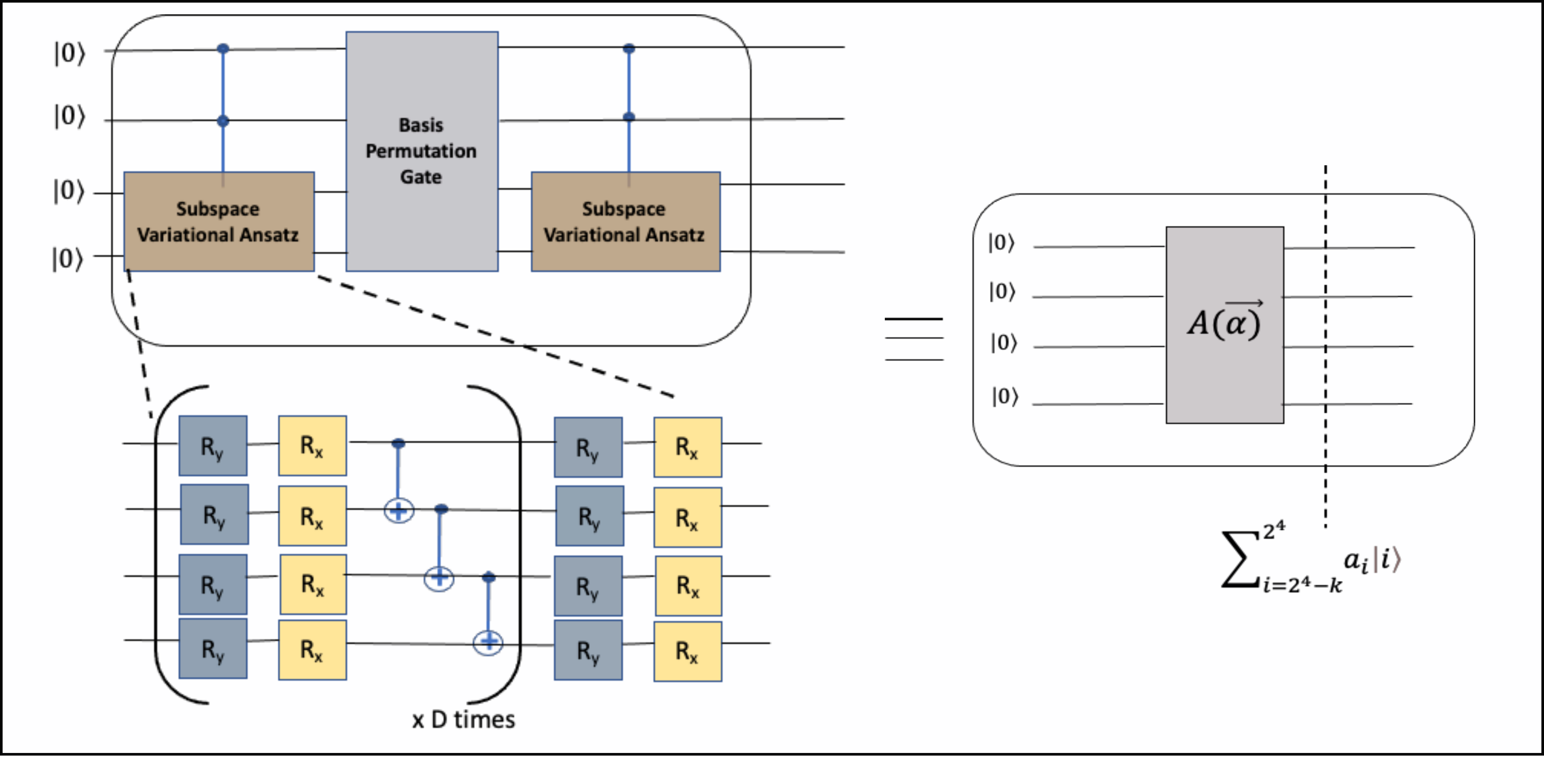}
\caption{Ansatz ($A(\alpha)$) that has been used to build the state within a k-dimensional space in Eq. \ref{A_ansatz}}
\label{fig:A_ansatz_circ}
\end{figure}

We start by first constructing a generic $k$-dimensional variational ansatz $A(\vec{\alpha})|0\rangle^{\bigotimes n}$ which allows us to subsequently confine the acceptable state within the symmetry subspace $\Omega$ defined in Eq.\ref{gen_opt_sch}, where $A(\vec{\alpha})$ is the unitary creating the variational ansatz with $\vec{\alpha}$ being the corresponding parameters. If $k$ is expressible as an exponentiation of 2, one could condition the variational circuit over the first $n-m$ qubits, where $m=log_2(k)$. Otherwise we choose $m = \lfloor log_2(k)\rfloor$ and then use a permutation gate to swap in the left over basis states before applying another layer of variational ansatz. Mathematically the action of the unitary $A(\vec{\alpha})$ can be envisioned as \begin{equation} \label{A_ansatz}
    A(\vec{\alpha})|0\rangle^{\bigotimes n} = \sum_{i=1}^{2^n} a_i(\vec{\alpha}) |i\rangle = \sum_{i=2^n -k}^{2^n} a_i(\vec{\alpha}) |i\rangle
\end{equation}
wherein $|i\rangle$ denotes the computational basis states and $a_i(\vec{\alpha})$ are the respective coefficients parameterized over $\vec{\alpha}$. The last equality in Eq. \ref{A_ansatz} follows from the fact that $a_i(\vec{\alpha}) =0 \:\:\forall\:\: i\:\:\le 2^n -k$. This allows us to create a variational ansatz over a $k$-dimensional subspace of $\mathbb{C}^{2^n}$ as is required.  Fig \ref{fig:A_ansatz_circ} provides a schematic view of the circuit involved in construction of the ansatz $A(\alpha)$. We now present two different methods that differs in the post-processing of the ansatz $A(\vec{\alpha})|0\rangle^{\bigotimes n}$ once created. As mentioned before, the ultimate goal of both the method will be to map the computational basis states $\{|i\rangle\}_{i=2^n -k}^{2^n}$ onto the symmetry sub-space $\Omega$ using a unitary transformation and then obtain the minimal energy eigenstate in that subspace.
%Fig \ref{fig:Circuit_Full} provides  schematic representation of the structure of the overall ansatz used in both the cases where the unitary $U(\vec{\theta})$ has been constructed using different methods. The Subspace variational Ansatz is constructed using a generalized scheme as shown in Fig \ref{fig:ansatz}

\subsection{Method 1- Exact Unitary}
This method uses an exact unitary that is composed of the eigenvectors of the symmetry operator. Let $U$ be the unitary operator that diagonalizes $O_k$ i.e. $U = \sum_{i=1}^{2^n} |\phi_i\rangle \langle i|$ wherein
$|\phi_i\rangle$ are the eigenvectors of the symmetry operator $O_k$. Using the unitary $U$ on the k-dimensional ansatz defined in Eq.\ref{A_ansatz},  we thus get,

\begin{equation} \label{U_methd_1}
    U A(\vec{\alpha})\ket{0}^{\bigotimes n} =  \sum_{j=1}^{2^n} |\phi_j\rangle \langle j|(\sum_{i=2^n-k}^{2^n} a_i(\vec{\alpha})\ket{i}) = \sum_{i=2^n-k}^{2^n} a_i(\vec{\alpha})\ket{\phi_i}
\end{equation}
where in the last equation in Eq.\ref{U_methd_1} only the eigenvectors {$|\phi_i\rangle$}$_{i=2^n-k}^{2^n}$ $\in \Omega$ survives. Operationally the matrix $U$ is constructed by stacking the $k$ eigenvectors corresponding to eigenvalue $S$ in the last $k$-columns. Note that the coefficients $a_i$ in Eq.\ref{U_methd_1} are explicitly dependant on tunable parameters $\vec{\alpha}$ imparted from the ansatz $A(\vec{\alpha})$. We thereafter train these parameters $\vec{\alpha}$ by minimizing the energy of the output state using the hamiltonian $H$ of the system as follows,

\begin{equation}
\label{eqn: ham_methd_1}
    \vec{\alpha}^{*} = argmin_{\vec{\alpha}}\:\: ^{n \bigotimes}\bra{0} A^\dagger(\vec{\alpha}) U^\dagger H U A(\vec{\alpha})\ket{0}^{\bigotimes n} 
\end{equation}

\begin{figure}[H]
\centering
\includegraphics[width=.75\textwidth]{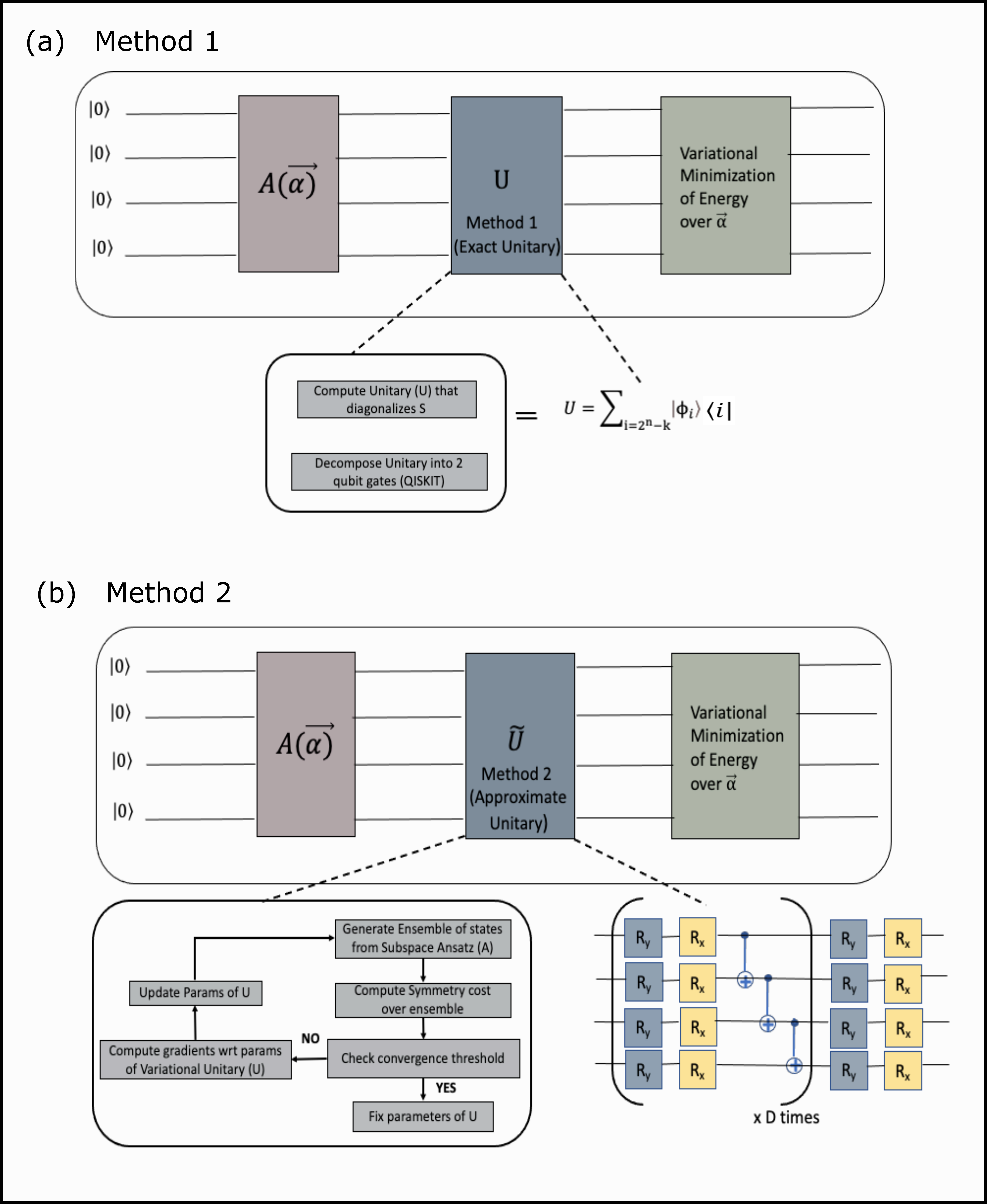}
\caption{The circuit decomposition and essential steps in (a) Method 1 and (b) Method 2}
\label{fig:U1_U2_circ}
\end{figure}

The minimization scheme in Eq.\ref{eqn: ham_methd_1} leads us to the minimal energy state within the symmetry subspace as is required. As this method uses the exact unitary $U$, it shall work irrespective of any initial ansatz with an output that is restricted to the subspace. We describe schematically method 1 in Fig. \ref {fig:U1_U2_circ}(a). This method however requires decomposing the matrix $U$ into a sequence of elementary gates which depending on the symmetry may result in a quantum circuit with very large depth. To circumvent this drawback, we propose an alternative approach (Method 2) described next.

\subsection{Method 2- Approximate Unitary construction}

%to approximately a unitary that suffices to achieve the minimally required mapping.

\begin{figure}[H]
\centering
\includegraphics[width=.75\textwidth]{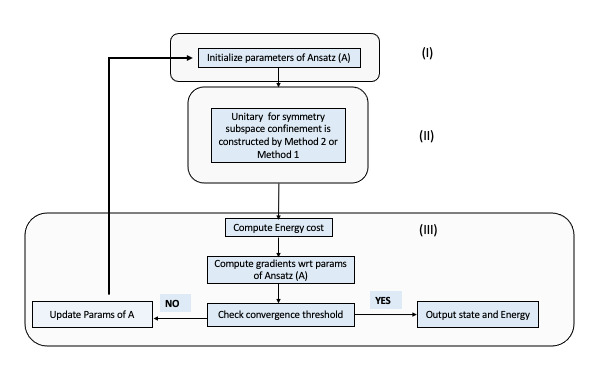}
\caption{Flowchart indicating the steps involved in the algorithm . (I) Indicates preparation of the ansatz as illustrated in Fig. \ref{fig:A_ansatz_circ}. (II) Preparation of the unitary that dictates symmetry restriction as discussed in Fig.\ref{fig:U1_U2_circ}. (III) The key steps in the variational optimization of the energy.}
\label{fig:flowchart}
\end{figure}

In this method, instead of using the exact unitary $U$ obtained from the diagonalization of the symmetry operator $O_k$ as introduced in Method 1, we approximate it using a parameterized ansatz $\tilde{U}$ which allows for a low depth quantum circuit construction. Operationally we use the same ansatz as in $A(\vec{\alpha})$ for constructing $\tilde{U}(\vec{\theta})$ and then learn the parameters variationally by minimizing the cost function $\langle(O_k-S)^2\rangle$ where $S$ is the desired eigenvalue. The state on which the unitary $\tilde{U}(\vec{\theta})$ acts is $A(\vec{\alpha})|0\rangle^{\bigotimes n}$ as introduced in the previous section. The variational minimization over $\vec{\theta}$ is done over many realization of the parameter set $\vec{\alpha}$ which is akin to an averaging procedure over the underlying sampling distribution $p(\vec{\alpha})$. Mathematically the parameter set $\vec{\theta}$ is learned as follows:
\begin{equation} \label{U_methd_2}
    \vec{\theta}^{*} =  argmin_{\vec{\theta}}\:\: \langle ^{n \bigotimes}\bra{0} A^\dagger(\vec{\alpha}) \tilde{U}^\dagger(\vec{\theta}) (O_k-S)^2
    \tilde{U}(\vec{\theta}) A(\vec{\alpha})\ket{0}^{\bigotimes n}\rangle_{p(\vec{\alpha})}
\end{equation}
where $\langle  \rangle_{p(\vec{\alpha})}$ represents the averaging over the distribution $p(\vec{\alpha})$. The parameters $\vec{\theta}^{*}$ are trained so as to achieve a very low margin of error of allowing $\tilde{U}$ to faithfully mimic the exact unitary $U$ and confine any subsequent operation to the symmetry subspace $\Omega$ irrespective of the state prepared by the ansatz $A$. With the parameters $\vec{\theta}$ known, one can proceed towards doing a variational optimization on the parameters $\vec{\alpha}$ just as before so as to minimize the energy according to the 
\begin{equation}
    \vec{\alpha}^{*} =  argmin_{\vec{\alpha}}\:\: ^{n \bigotimes}\bra{0} A^\dagger(\vec{\alpha}) \tilde{U}^\dagger(\vec{\theta^*}) H
    \tilde{U}(\vec{\theta^*}) A(\vec{\alpha})\ket{0}^{\bigotimes n}
\end{equation}
We describe schematically method 2 in Fig. \ref {fig:U1_U2_circ}(b).
In Fig.\ref{fig:flowchart} we describe entire algorithm that is followed for both method 1 and method 2 and highlight the essential steps. Since we have discussed in detail the first two key steps of the algorithm which includes preparation of the ansatz using $A(\vec{\alpha})\ket{0}^{\bigotimes n}$ (see Fig.\ref{fig:A_ansatz_circ}) and then construction and operation by the unitary which confines the state within a symmetry subspace (see Fig.\ref{fig:U1_U2_circ}), we emphasize in somewhat detail the subsequent steps which articulates the procedure for obtaining the ground state energy in Fig.\ref{fig:flowchart} (designated as (III)). The procedure is similar to any variational hybrid algorithms routinely used nowadays. One computes the average energy using the Hamiltonian of the system and the parameterized state constrained in the symmetry sub-space (from step II in Fig.\ref{fig:flowchart}). The gradient of this energy with respect to the parameters of the state is constructed and the convergence of the norm of the gradient is checked. If the desired threshold is not attained, parameters of the state is further changed (by changing $\vec{\alpha}$ in step I in Fig. \ref{fig:flowchart}) for the next iteration. The unitaries necessary to accomplish this operation comes from the standard procedure of the conversion of the system Hamiltonian into Pauli-strings. The accompanying circuit description of such unitaries would then be highly system specific

%%%%%%%%%%%%%%%%%%%%%%%%%%%%%%%%%%%%%%%%%%
\section{Results}
Both methods discussed above have been tested against $XXZ$ spin Hamiltonian and $H_2$ molecule hamiltonian. We plot against chosen symmetries, the energy error and state fidelity against the low energy state that respects the symmetry. For method 2 that makes use of an approximate Unitary construction, we show in addition, the deviation in the expected symmetry of output state from the chosen value. The results have been obtained using Qiskit \cite{Qiskit} state-vector simulator. The parameter updates in every iteration has been globally bound by 0.5 radians and reduced iteratively for convergence. The use of only $R_y$ and $R_x$ gates in our ansatz allows for gradients to be calculated with a shift of $\pi$ radians on the respective gate parameters. 

\subsection{$XXZ$ Spin Hamiltonian}

The $XXZ$ spin hamiltonian is given by,
\begin{equation}
    H = \sum_{i} J(\sigma^x_i\sigma^x_{i+1} + \sigma^y_i\sigma^y_{i+1}) + K\sigma^z_i\sigma^z_{i+1}
\end{equation}
where the summation is over the nearest neighbour spins with open boundaries. We study the low energy states respected by the Reflection and Rotation Symmetry \cite{spin_half}. 

Both symmetry operators have eigenvalues of $\pm 1$. We would like to variationally probe the low energy states in each of these subspaces. We set the number of spins to be $4$ and work with $J=1$ and $K=3$. This corresponds to $16$ dimensional Hilbert space. For method 2 we have made use of D=5 layers as shown in Fig \ref{fig:U1_U2_circ} to train the unitary $\tilde{U}(\vec{\theta})$ up to a mean error of 0.001 on the Symmetry value  for 100 random samples generated by the ansatz $A(\vec{\alpha})$

\subsubsection{Reflection Symmetry}

\begin{figure}[H]
\centering
\subfigure[Method 1: Energy error vs iterations]{
\includegraphics[width=.35\textwidth]{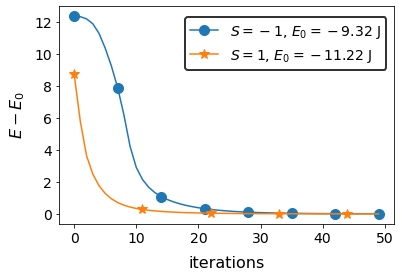}
}
\subfigure[Method 1: State fidelity vs iterations]{
\includegraphics[width=.35\textwidth]{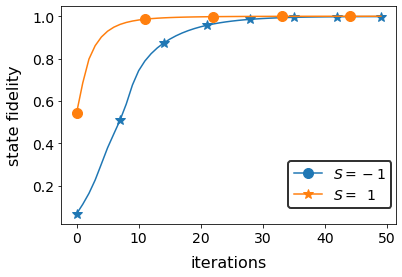}
}
\subfigure[Method 2: Energy error vs iterations]{
\includegraphics[width=.35\textwidth]{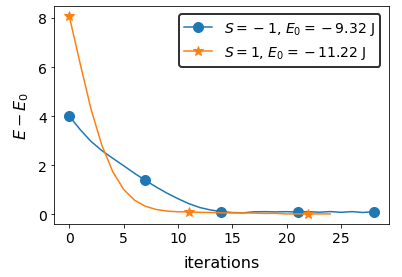}
}
\subfigure[Method 2: State fidelity vs iterations]{
\includegraphics[width=.35\textwidth]{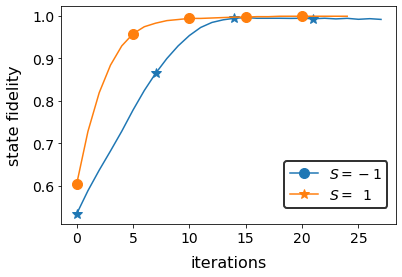}
}
\subfigure[Method 2: Symmetry value vs iterations]{
\includegraphics[width=.35\textwidth]{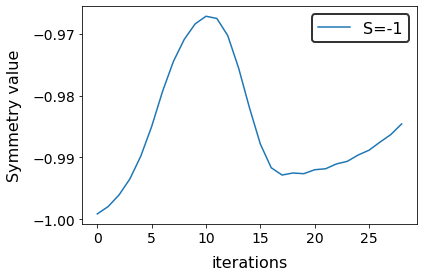}
}
\subfigure[Method 2: Symmetry value vs iterations]{
\includegraphics[width=.35\textwidth]{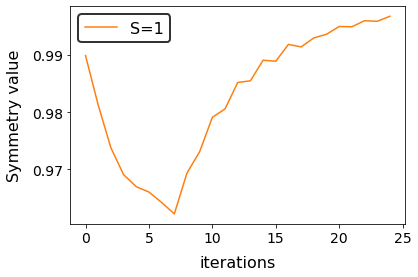}
}
\caption{Computing the lowest energy state of XXZ Hamiltonian within subspace labelled by the Reflection Symmetry Operator using Unitary constructed from method 1 and method 2}
\label{fig:Reflection_Exact_Unitary}
\end{figure}

The reflection symmetry operator is given by, $S = \Pi_i(\sigma^x_i\sigma^x_{i+1} + \sigma^y_i\sigma^y_{i+1} +\sigma^z_i\sigma^z_{i+1} + I)/2$. The reflection operator has a 6 dimensional subspace that measures -1 and 10 dimensional subspace that measures +1. Fig \ref{fig:Reflection_Exact_Unitary} shows energy error and state fidelity plots indicating the convergence of the variational methods to the exact solution for each of the Symmetry values. Note that in method 2, for $S=-1$, the method converges at the correct solution, despite the overall ground state energy being much lower. We notice that the symmetry value fluctuates within a very small interval around the exact value during the training.

\subsubsection{Rotation Symmetry}
\begin{figure}[H]
\centering
\subfigure[Method 1: Energy error vs iterations]{
\includegraphics[width=.35\textwidth]{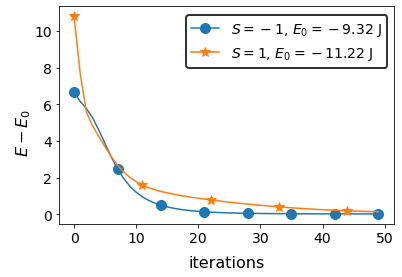}
}
\subfigure[Method 1: State fidelity vs iterations]{
\includegraphics[width=.35\textwidth]{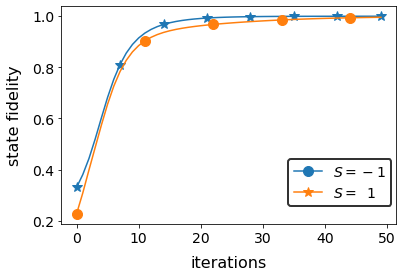}
}
\subfigure[Method 2: Energy error vs iterations]{
\includegraphics[width=.35\textwidth]{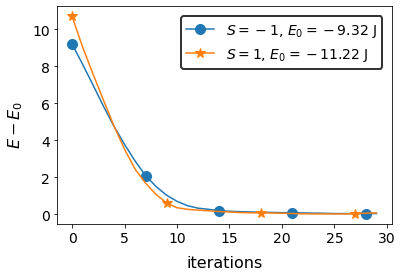}
}
\subfigure[Method 2: State fidelity vs iterations]{
\includegraphics[width=.35\textwidth]{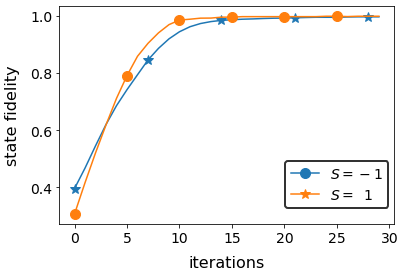}
}
\subfigure[Method 2: Symmetry value vs iterations]{
\includegraphics[width=.35\textwidth]{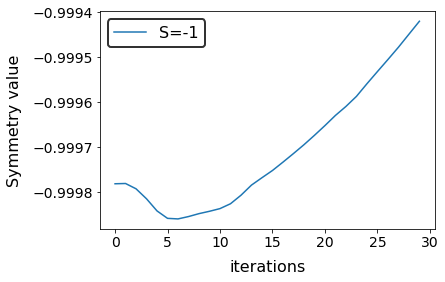}
}
\subfigure[Method 2: Symmetry value vs iterations]{
\includegraphics[width=.35\textwidth]{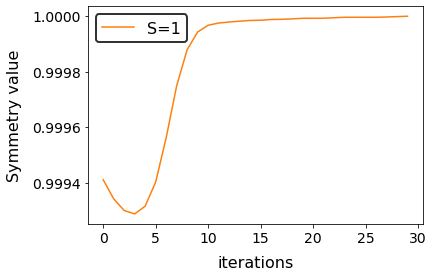}
}
\caption{Computing the lowest energy state of XXZ Hamiltonian within subspace labelled by the Rotation Symmetry Operator using Unitary constructed from method 1 and method 2}
\label{fig:Rotation_Exact_Unitary}
\end{figure}

The reflection symmetry operator is given by, $S = \Pi_i(\sigma^x_i)$. The symmetry subspace of the rotation operator with eigenvalue $\pm 1$ is 8 dimensional. Fig \ref{fig:Rotation_Exact_Unitary} shows energy error and state fidelity plots indicating the convergence of the variational methods to the exact solution for each of the Symmetry values. Note that in method 2, for $S=-1$, the method converges at the right solution, despite the overall ground state energy being much lower. We notice that the symmetry value fluctuates within a very small interval around the exact value during the training.

\subsection{$H_2$ Hamiltonian}

\begin{figure}[H]
\centering
\subfigure[Method 1: Energy error vs iterations]{
\includegraphics[width=.35\textwidth]{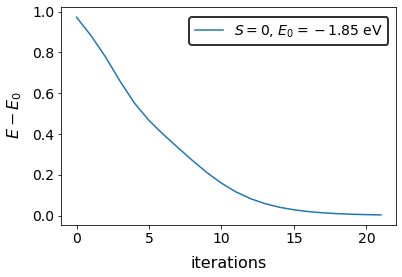}
}
\subfigure[Method 1: State fidelity vs iterations]{
\includegraphics[width=.35\textwidth]{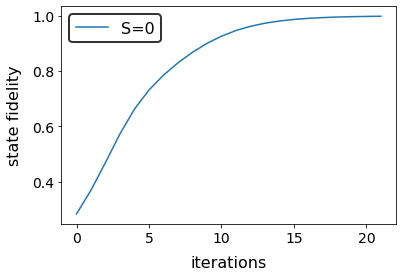}
}
\subfigure[Method 2: Energy error vs iterations]{
\includegraphics[width=.35\textwidth]{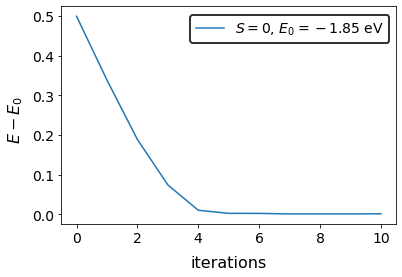}
}
\subfigure[Method 2: State fidelity vs iterations]{
\includegraphics[width=.35\textwidth]{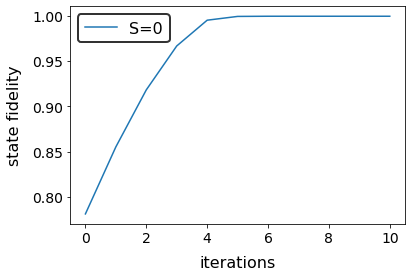}
}
\subfigure[Method 2: Symmetry value vs iterations]{
\includegraphics[width=.35\textwidth]{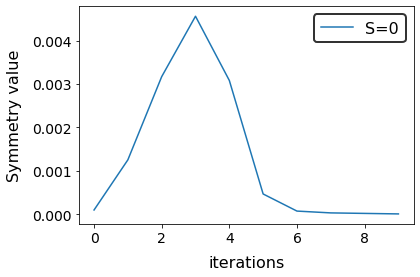}

}
\caption{Computing the lowest energy state of the hydrogen Hamiltonian within $\sum_i S^i_z=0$ subspace labelled by  $S^2=s(s+1)=0$ using Unitary constructed from method 1 and method 2}
\label{fig:Hydrogen}
\end{figure}

As an example of a real-molecular system we choose the prototypical $H_2$ molecule. The electronic Hamiltonian for the system is constructed in the $STO-3G$ basis in the subspace spanned by vectors of the form |$n_2(\downarrow), n_1(\downarrow), n_2(\uparrow), n_1(\uparrow)\rangle$ with $\sum_i S_z^i =0$ and $N=\sum_i a^\dagger_i a_i =2$ (the four such states are 
$|0110\rangle, |0101\rangle, |1010\rangle, |1001\rangle$). The corresponding matrix (in units of eV) at the equilibrium bond length of 0.725 \AA $\:$ is \cite{2021-h2}:
\begin{equation}
H=\begin{pmatrix}
-1.06 & 0 & 0 & 0.18\\
0 & -1.84 & 0.18 & 0 \\
0 & 0.18 & -0.23 & 0 \\
0.18 & 0 & 0 & -1.06
\end{pmatrix}
\end{equation}
Since the entire space is spanned by states with the zero eigenvalue of $\sum_i S_z^i$, we have used the total angular momentum squared as the symmetry operator of our choice ($O_k = \hat{S}^2$). The latter operator in the aforesaid basis
\begin{equation}
S^2=\frac{1}{2}\begin{pmatrix}
1.0 & 0 & 0 & -1.0\\
0 & 0 & 0 & 0 \\
0 & 0 & 0 & 0 \\
-1.0 & 0 & 0 & 1.0
\end{pmatrix}
\end{equation}
Only one of the eigenvalues of the $S^2=s(s+1)$ matrix is 1 and the remaining ones are all 0. The techniques developed in this report  requires the symmetry subspace to be of dimension greater than 1. Thus we shall be restricting to the subspace $S^2=s(s+1)=0$.  
For method 2 we have made use of D=4 layers as shown in Fig \ref{fig:U1_U2_circ} to train the unitary $U(\vec{\theta})$ up to a mean error of 0.001 on the Symmetry value  for 100 random samples generated by the ansatz $A(\vec{\alpha})$.  Fig \ref{fig:Hydrogen} shows energy error and state fidelity plots indicating the convergence of the variational methods to the exact solution for $S^2=s(s+1)=0$. Here again, we notice that the symmetry value fluctuates within a very small interval around the exact value during the training.

%%%%%%%%%%%%%%%%%%%%%%%%%%%%%%%%%%%%%%%%%%
\section{Discussion}
We discuss two general methods of restricting the exploration of a symmetry labelled subspace for a variational ansatz in identifying the ground state energy of a spin $XXZ$ Hamiltonian and $H_2$ Hamiltonian. Method 1 makes use of an exact Unitary constructed from the Symmetry operator giving raise to a large circuit decomposition. This is overcome through the use of variationally trained Unitaries developed in method 2. The methods developed in here work more generally where the constraint function need not commute with Hamiltonian,i.e, need not be a symmetry, as reflected in construction of the unitary $\tilde{U}(\vec{\theta})$. We would also like to point out that method 2 provides for a useful tool in developing Unitaries variationally that satisfy a property of interest. The depth of the circuit and thus the number of parameters to be trained is likely to increase with the number of qubits. As most symmetry operators of interest are usually defined with the similar operators acting over the entire qubit space, as a future work one might want to investigate about minimal 2 qubit unitaries  that leave the Symmetry value unchanged and use them to create generalized ansatz over several qubits.

%%%%%%%%%%%%%%%%%%%%%%%%%%%%%%%%%%%%%%%%%%
%\section{Conclusions}

%Maybe shift the future work to here

%%%%%%%%%%%%%%%%%%%%%%%%%%%%%%%%%%%%%%%%%%
\vspace{6pt} 

%%%%%%%%%%%%%%%%%%%%%%%%%%%%%%%%%%%%%%%%%%
%% optional
%\supplementary{The following are available online at \linksupplementary{s1}, Figure S1: title, Table S1: title, Video S1: title.}


%%%%%%%%%%%%%%%%%%%%%%%%%%%%%%%%%%%%%%%%%%
%\authorcontributions{Conceptualization, X.X. and Y.Y.; methodology, X.X.; software, X.X.; validation, X.X., Y.Y. and Z.Z.; formal analysis, X.X.; investigation, X.X.; resources, X.X.; data curation, X.X.; writing---original draft preparation, X.X.; writing---review and editing, X.X.; visualization, X.X.; supervision, X.X.; project administration, X.X.; funding acquisition, Y.Y. All authors have read and agreed to the published version of the manuscript.}

\bibliographystyle{unsrt}
\bibliography{output}

\end{document}